\documentclass[twocolumn]{raa}

\usepackage{epsfig,amsmath,enumerate,amssymb,aas_macros,bm,booktabs,multirow,xcolor}
\usepackage{graphics}
\usepackage{multirow}
\usepackage{natbib}
\usepackage{times}
\usepackage{epsfig}
\bibpunct{(}{)}{;}{a}{}{,}

\begin{document}
\title{The effect of hydrodynamics alone on the subhalo population in a $\Lambda$CDM rich cluster sized dark matter halo}

 \volnopage{ {\bf 20XX} Vol.\ {\bf X} No. {\bf XX}, 000--000}
   \setcounter{page}{1}
   
    \author{Junyi Jia\inst{1,2}, Liang Gao\inst{1,2,3}, Yan Qu\inst{1}}

   \institute {Key Laboratory for Computational Astrophysics, National Astronomical Observatories, Chinese Academy of Sciences, Beijing, 100101 {\it jyjia@nao.cas.cn}\\
     \and
      School of Astronomy and Space Science, University of Chinese Academy of Sciences, Beijing 100049\\
\and 
	Institute of Computational Cosmology, Department of Physics, University of Durham, South Road, Durham DH1 3LE, UK\\
\vs \no
   {\small Received 20XX Month Day; accepted 20XX Month Day}
}

\abstract{We perform a set of non-radiative hydro-dynamical (NHD) simulations of a rich cluster sized dark matter halo from the Phoenix project with 3 different numerical resolutions, to investigate the effect of hydrodynamics alone on the subhalo population in the halo. Compared to dark matter only (DMO) simulations of the same halo, subhaloes are less abundant for relatively massive subhaloes ($M_{sub} > 2.5 \times 10^9h^{-1}M_{\odot}$, or $V_{max} > 70 kms^{-1}$) but more abundant for less massive subhaloes in the NHD simulations. This results in different shapes in the subhalo mass/$V_{max}$ function in two different sets of simulations. At given subhalo mass, the subhaloes less massive than $10^{10}  h^{-1}M_{\odot}$ have larger $V_{max}$ in the NHD than DMO simulations, while $V_{max}$ is similar for the subhaloes more massive than the mass value. This is mainly because the progenitors of present day low mass subhaloes have larger concentration parameters in the NHD than DMO simulations. The survival number fraction of the accreted low mass progenitors of the main halo at redshift 2 is about 50 percent higher in the NHD than DMO simulations.
}

   \authorrunning{Junyi Jia, et al.}            
   \titlerunning{The effect of hydrodynamics alone on the subhalo population in a $\Lambda$CDM rich Cluster}  
   \maketitle

\section{INTRODUCTION}
\label{sec:intro}

In the standard $\Lambda$CDM Cosmology, structure formation is hierarchical. Small dark matter haloes formed firstly, and then merge to form larger and larger systems. During this hierarchical clustering process, earlier accreted halo often survive as a subhalo to orbit its host. As the most massive subhaloes are expected to be the hosts of luminous satellite galaxies, properties of these subhaloes have been extensively investigated in the past two decades (e.g. \citealt{Moore1998}, \citealt{Ghigna1998}, \citealt{Moore1999}, \citealt{Klypin1999},  \citealt{Ghigna2000}, \citealt{Springel2001a}, \citealt{Stoehr2002}, \citealt{Stoehr2003}, \citealt{De Lucia2004}, \citealt{Diemand2004}, \citealt{Gao2004}, \citealt{Kravtsov2004}, \citealt{Gao2012}, \citealt{Han2016}, \citealt{Han2018}, \citealt{vandenBosh2018}).

Limited by computational power, earlier works on subhaloes have been confined to use high resolution DMO  simulations, whilst baryonic physics may play a sizeable impact on the subhalo population in dark matter haloes. Thanks to great advance in super computer power in recent years, properties of subhaloes have been widely studied with modern hydro-dynamical simulations with galaxy formation models (e.g. \citealt{Duffy2010},  \citealt{Libeskind2010}, \citealt{Romano-Diaz2010}, \citealt{Garrison-Kimmel2017}, \citealt{Graus2018}, \citealt{Richings2020}). Compared to DMO simulations,  in general, the subhalo population is found to be less abundant in hydrodynamical simulations. However, the degree of the difference in the subhalo population between the two varies with the adopted different galaxy formation models, as structure of dark matter halo/subhalo is sensitive to the galaxy formation models. For examples, \cite{Nagai2005}, \cite{Maccio2006} and \cite{Weinberg2008} find that baryon has only a small impact on the subhalo population; \cite{Sawala2013}, \cite{Schaller2015} and \cite{Sawala2017} show that the impact of baryonic physics depends on the subhalo mass, above $10^{12}M_{\odot}-10^{13}M_{\odot}$ the subhalo abundance ratio between the hydrodynamical and DMO runs is close to 1, but it drops below 1 for low mass subhaloes. \cite{Zhu2016} report a similar results in a set of simulations of Milky Way-size dark matter haloes.  \cite{Chua2017} claims that baryonic physics changes the shape of the subhalo mass function: relative to the DMO simulations, subhalo is less abundant in the low-mass end and more abundant for relatively larger subhaoles in their hydrodynamical simulations. 

In this paper, we compensate these studies by investigating the impact of hydrodynamics alone on the subhalo population in a $\Lambda$CDM rich cluster sized halo. Compared to earlier works on this topic(e.g., \citealt{Lin2006}), our NHD simulations have much higher mass and force resolutions, and have 3 different numerical resolutions. The latter allows us to carry out numerical convergence study. Our paper is organized as follows. In section 2, we introduce the simulation sets used in this study.  Section 3 compares the subhalo abundance, internal structure and the evolution of subhaloes in different sets of simulations. We draw our conclusin in Section 4. 

\section{THE SIMULATIONS}
\label{sec:simu}

\begin{table*}
	\centering    
    \caption{Basic parameters of simulations. Each of the simulation is labelled as Ph-A-N(-g), where N identifies numerical resolution level from 2-4 (2 is the highest), and g represent simulation with gas. $M_{DM}$ and $M_{gas}$ represent dark matter particle mass and gas particle mass in the high-resolution region that includes the cluster respectively; $M_{200}$ and $R_{200}$ are the virial mass and virial radius of the halo respectively. The parameter $\epsilon$ is gravitational softening length.}
	\label{tab:simu_table}
	\begin{tabular}{|c|c|c|c|c|c|} 
		\hline
		Name & $M_{DM}$ & $M_{gas}$ & $M_{200}$ & $R_{200}$ & $\epsilon$ \\
         & [$h^{-1}M_{\odot}$] & [$h^{-1}M_{\odot}$] & [$h^{-1}M_{\odot}$] & [$h^{-1}Mpc$] & [$h^{-1}kpc$]\\
		\hline
		Ph-A-2 & $5.084*10^{6}$ & --- & $6.596*10^{14}$ & 1.416 & 0.32 \\
		Ph-A-3 & $1.716*10^{7}$ & --- & $6.599*10^{14}$ & 1.416 & 0.7 \\
		Ph-A-4 & $1.373*10^{8}$  & --- & $6.598*10^{14}$ & 1.416 & 2.8 \\
		Ph-A-2-g & $4.271*10^{6}$ & $8.134*10^{5}$ & $6.680*10^{14}$ & 1.422 & 0.32\\
		Ph-A-3-g & $1.441*10^{7}$ & $2.746*10^{6} $ & $6.664*10^{14}$ & 1.420 & 0.7\\
		Ph-A-4-g & $1.153*10^{8}$ & $2.197*10^{7}$ & $6.676*10^{14}$ & 1.421 & 2.8\\
		\hline
	\end{tabular}
\end{table*}

	\begin{figure*}
	\centering
	\includegraphics[width=\textwidth]{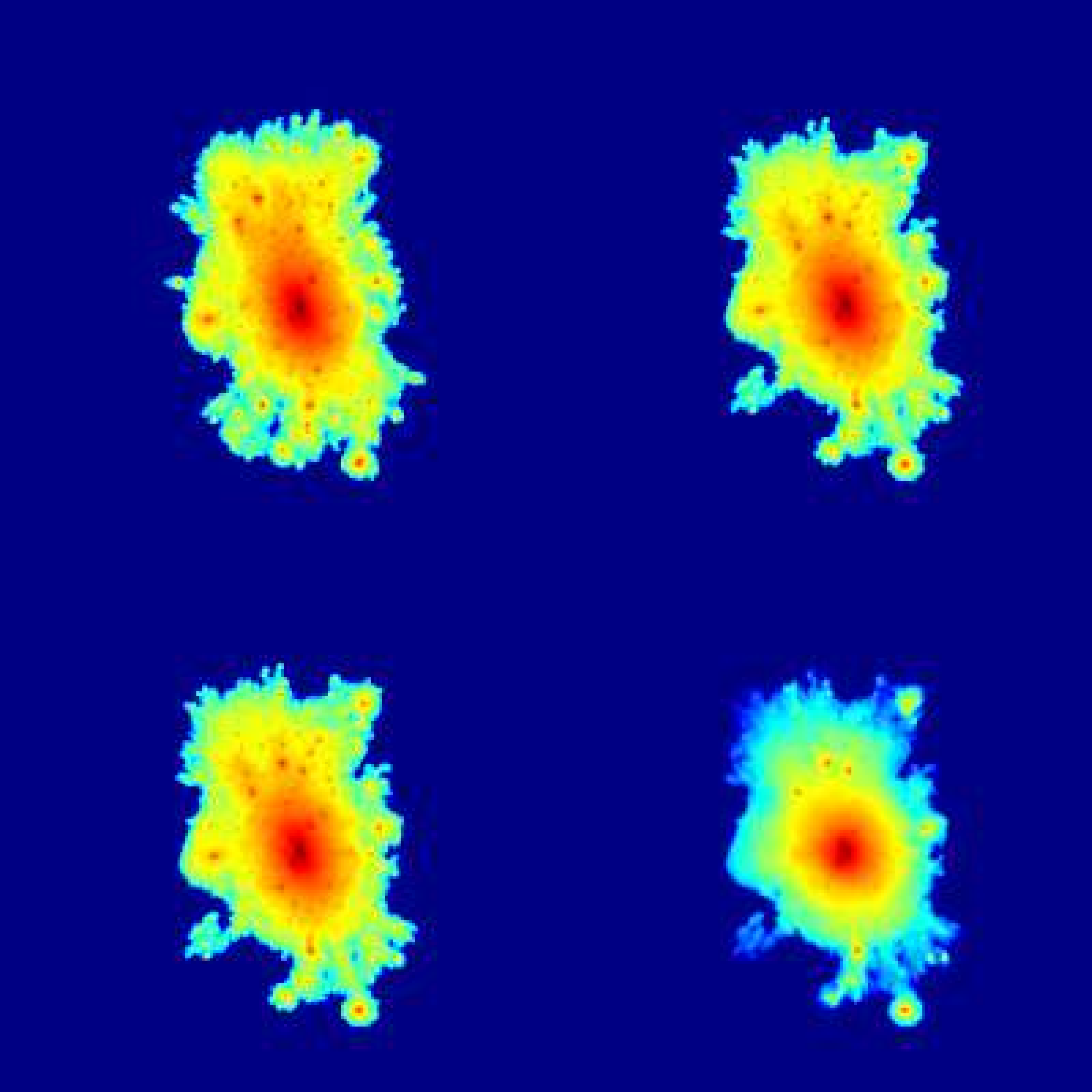}
    \caption{Density maps of the Ph-A FOF group at $z=0$. Top panels show the total density map of the Ph-A-2 FOF group, the left for the Ph-A-2 run and the right for the Ph-A-2-g run. Bottom panels show the dark matter (left panel) and the gas density map of the FOF group in the Ph-A-2-g simulation.}
	\label{fig:slice}
	\end{figure*} 	
    
The numerical simulations used in this work comprise two sets of ultra-high resolution re-simulation of a cluster-sized dark matter halo and its surrounding. This halo is selected from the Phoenix Project and is termed as the Ph-A halo. The Phoenix Project resimulated the formation and evolution of 9 different cluster-sized dark matter haloes selected from the Millennium Simulation (\citealt{Springel2005}) at ultra-high resolution.  We refer readers to \citet{Gao2012} for details of the Phoenix Project. The Ph-A halo has a virial mass $M_{200} \sim 6.6 \times 10^{14}h^{-1}M_{\odot}$ and a virial radius $R_{200} \sim 1.4 \times h^{-1}Mpc$ which is defined as the radius within which the enclosed mean density is 200 times the critical density of the Universe. In the Phoenix project the halo has been simulated at 4 different resolution levels, referred to as Ph-A-1, -2, -3 and -4 where 1 is the highest and 4 is the lowest level. In this study we re-run the Ph-A-2, -3 and -4 with the hydrodynamic code GADGET-3 (\citealt{Springel2005}) and denote them with a suffix ``-g’’ to indicate the inclusion of hydrodynamics. The highest resolution simulation, Ph-A-2-g, has a mass resolution of $4.3 \times 10^6 h^{-1}M_{\odot}$ for dark matter and  $8.1 \times 10^{5} h^{-1}M_{\odot}$ for gas particles. All the simulations in this work adopt the cosmological parameters, ${\Omega_{\rm m}}=0.25,  {\Omega_{\rm \Lambda}}=0.75, h=0.73, \sigma_{8}=0.9, n_s=1$, and ${\Omega_{\rm b}}=0.045$. The  detailed information of these simulations are summarized in Table~\ref{tab:simu_table}.

Dark matter haloes of the simulations are identified by standard friends-of-friends algorithm (\citealt{Davis1985}) with a linking length of 0.2 times mean interparticle separation. Based on dark matter halo catalogue, we further identify locally overdense and selfbound subhaloes with the SUBFIND algorithm (\citealt{Springel2001b}).  The merger trees of the dark matter haloes and subhaloes are constructed using the D-Trees algorithm (\citealt{Jiang2014}) which identifies the descendant of an object at next output time by tracing the most bound particles of the object.

In Fig.~\ref{fig:slice} , we present a visual impression of the Ph-A-2  and Ph-A-2-g FOF haloes at $z=0$. The upper panel compares the density distribution of total matter in the Ph-A-2 and Ph-A-2-g simulations. Note that only particles belonging to the FOF groups are used in making the projections. While the total matter distribution of the group is overall similar, some notable differences can still be found, massive subhaloes are more pronounced in the DMO simulation, and relatively central region is round in the Ph-A-2-g run. The lower panel shows separately the dark matter and the gas distribution in the Ph-A-2-g. Compared to the dark matter, the gas distribution is less clumpy, in agreement with many existing results(e.g., \citealt{Fang2009}).
	\begin{figure*}
	\includegraphics[width=\textwidth]{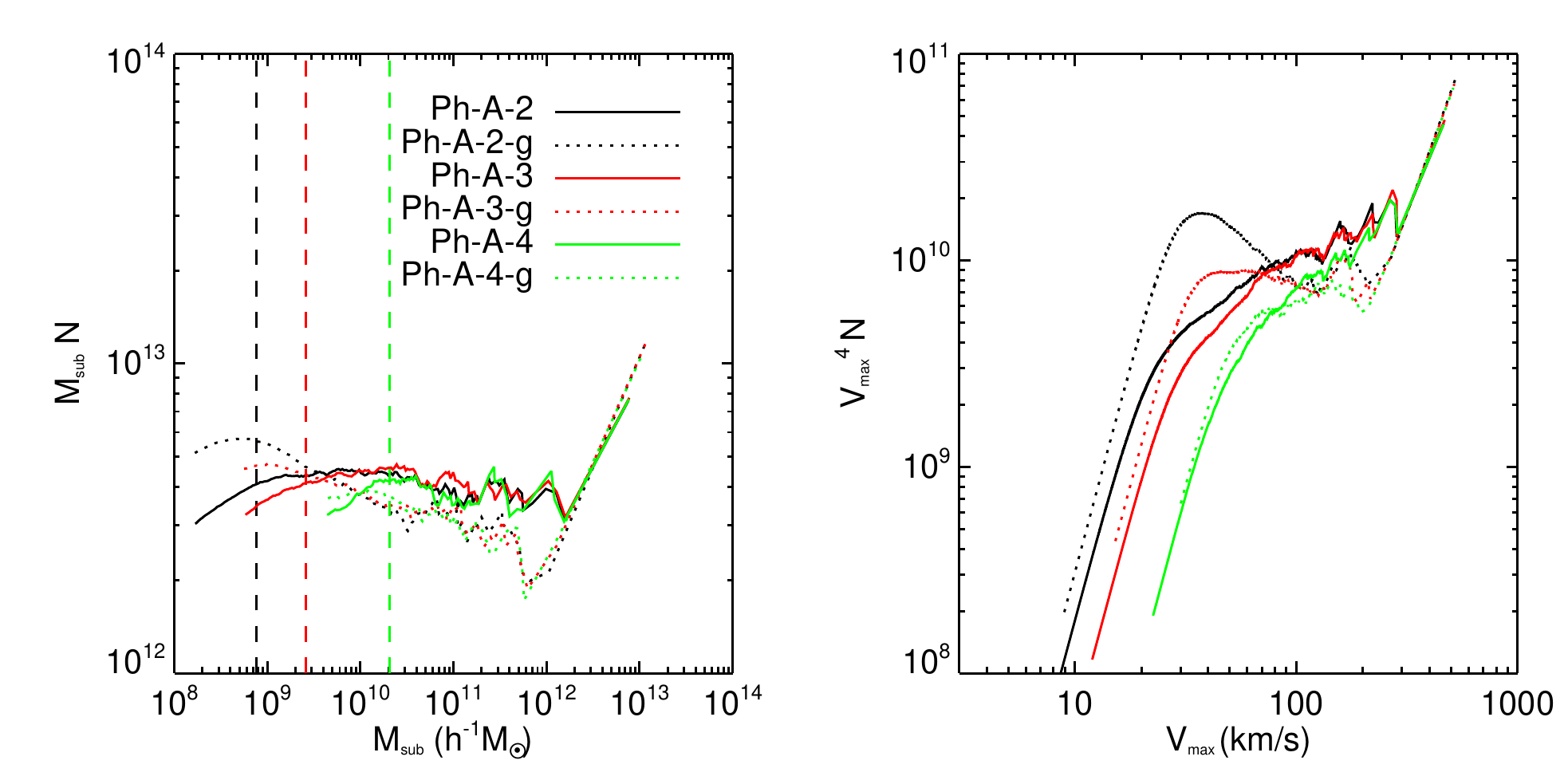}
    \caption{Left/right panel is the cumulative subhalo mass/$V_{max}$ functions of the Ph-A halo in two sets of simulations with three different resolutions at $z=0$. Note that the mass/$V_{max}$ functions shown here are multiplied by subhalo mass/$V_{max}$ in order to take out the dominated mass/$V_{max}$ dependence. The solid and dotted lines show results for the DMO and NHD simulations, respectively. Results from different resolution runs are distinguished with different colours as shown in the label. The vertical dotted lines show the masses of 150 dark matter particles in the DMO simulations at each resolution.}
	\label{fig:m&vmax_f}
	\end{figure*}

\section{Results}
\label{sec:stat}

\subsection{The subhalo mass and $V_{max}$ function}
\label{sec:mass_func}

 The left panel of Fig.~\ref{fig:m&vmax_f} shows the cumulative subhalo mass function for both DMO (solid lines) and NHD runs (dotted lines). Different colours are used to distinguish different resolutions as shown in the label. The cumulative subhalo mass functions are multiplied by subhalo mass in order to remove the dominant mass dependence. Apparently, massive  subhaloes are more abundant in the DMO than in the NHD run, in agreement with many existing studies (e.g., \citealt{Dolag2009}, \citealt{Schaller2015}, \citealt{Zhu2016}, \citealt{Sawala2017}), while the difference becomes smaller with decreasing subhalo mass. Below $2 \times 10^9h^{-1}M_{\odot}$, interestingly, the NHD runs contain more abundant subhaloes than their DMO counterparts. The other noticeable fact is that the slope of the cumulative subhalo mass function is $-1.1$, in the NHD runs, steeper than that of the DMO simulations (-1.0). Note that the cumulative subhalo mass functions between different resolution simulations converge down to subhaloes containing about 150 dark matter particles for both DMO and NHD simulations(vertical dotted line in left panel of Fig.~\ref{fig:m&vmax_f}).

The suppression in the subhalo mass function in hydrodynamic simulations of rich cluster sized dark matter haloes in the subhalo mass range $10^{10}h^{-1}M_{\odot}< M_{sub} < 10^{12}h^{-1}M_{\odot}$ is also found by \citet{Dolag2009}. The difference may be understood as the subhaloes in the NHD simulations suffer from additional ram pressure stripping in relative to the DMO runs.(\citealt{Tormen2004};\citealt{Puchwein2005}).

The cumulative $V_{max}$ functions weighted by subhalo $V_{max}^4$ are presented in the right panel of the Fig.\ref{fig:m&vmax_f}. In agreement with many existing results (e.g., \citealt{Zhu2016}), $V_{max}$ functions in the DMO runs have higher amplitude than their NHD counterparts for relatively large $V_{max}$ subhaloes, while below $V_{max} \sim 70 ~kms^{-1}$, the relation is reversed that the abundance of subhaloes in our NHD simulation exceeds the DMO runs. Also the slope of the $V_{max}$ function is much steeper in the NHD than DMO simulations, which is $-4.4$ for the NHD runs and $-3.4$ for the DMO runs, respectively.  The scale dependence of differences in the subhalo mass/$V_{max}$ functions may be caused by different physics playing at different scales, e.g. subhalo mass/$V_{max}$  dependence of interplay between tidal force and ram pressure as will be discussed later.

\subsection{The structure of subhaloes}

    \begin{figure*}
	\includegraphics[width=\textwidth]{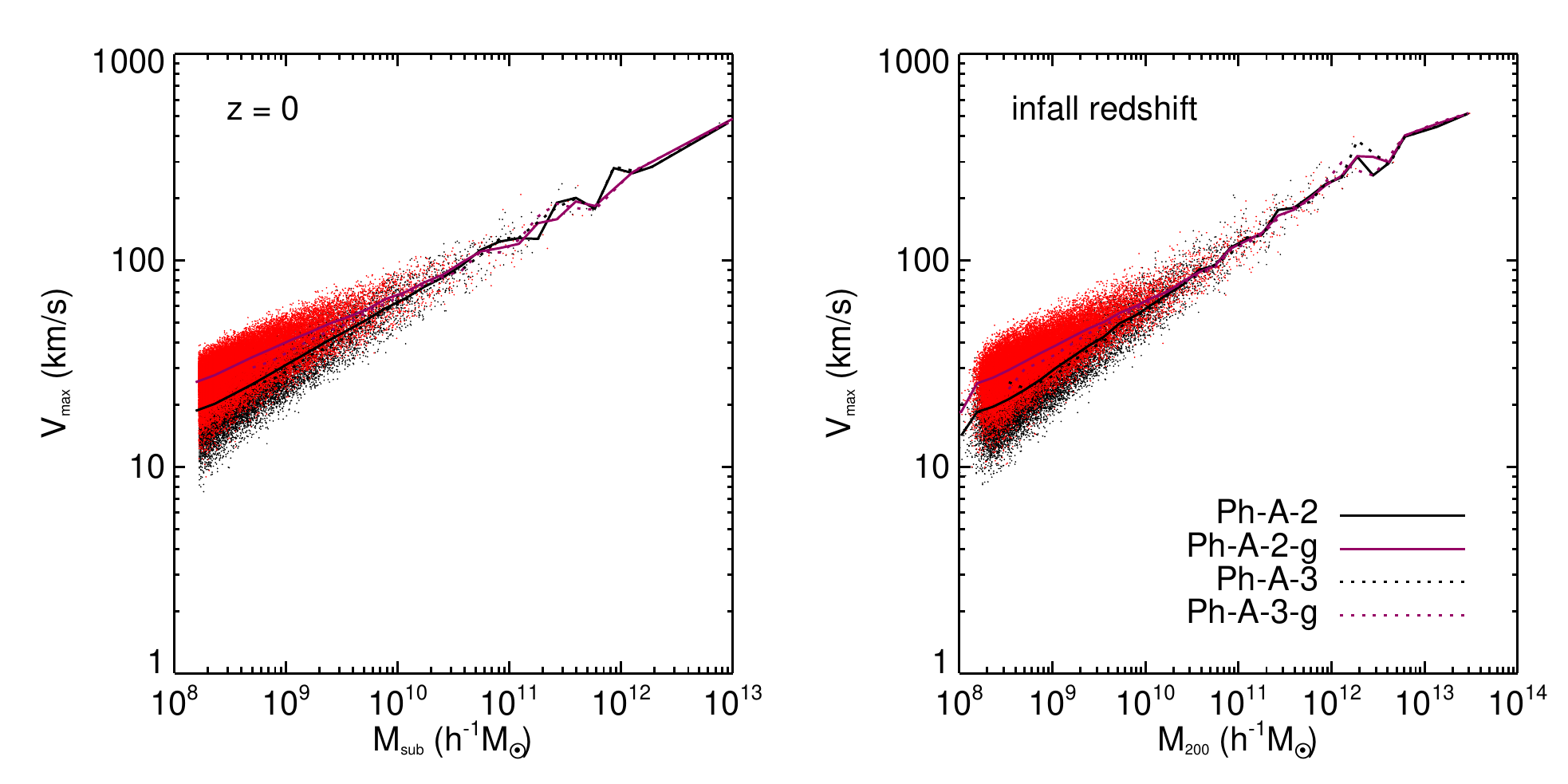}
    \caption{Left panel: $V_{max} - {\rm M_{sub}}$ relations for subhaloes in the level-3 and level-2 simulations at $z=0$ median values are shown. Red and black dots here represent subhalos in Ph-A-2 and Ph-A-2-g, respectively. Results of median values for the level 2 and 3 are distinguished with different line styles, and the DMO and NHD runs are distinguished with different colors are indicated in the label of the right panel. Right panel: $V_{max}$ - ${\rm M_{200}}$ relations for the progenitors of all present day subhaloes at infall.}
	\label{fig:m_vs_vmax_z0}
	\end{figure*}
    
    It is interesting to see how the hydrodynamics affect the internal structure of subhaloes. The left panel of Fig.~\ref{fig:m_vs_vmax_z0} shows $V_{max}-M_{sub}$ relation of subhaloes in both DMO and NHD simulations with the level-2 and level-3 resolutions at $z=0$, median values of $V_{max}$ in each mass bin are shown. The results for the level-2 and level-3 runs are shown with solid and dashed lines, respectively; and the results for the DMO and NHD runs are distinguished with different colors as shown in the label. Interestingly, for subhaloes more massive than about $2 \times 10^{10}h^{-1}M_{\odot}$, $V_{max}-M_{sub}$ relation is nearly identical between two sets of simulations, while below this mass, subhaloes tend to have higher $V_{max}$ in the NHD simulations. The difference becomes larger with decreasing subhalo mass. At $M_{sub} =10^9h^{-1}M_{\odot}$, the median value of $V_{max}$ of subhaloes in the NHD simulations is about $30$ percent larger than that of the DMO runs. These results suggest that low-mass subhaloes in the NHD runs are more concentrated than their DMO counterparts. The good agreements between different resolutions suggest that our results are robust to changes in numerical resolution.
   
   	\begin{figure*}
	\includegraphics[width=\textwidth]{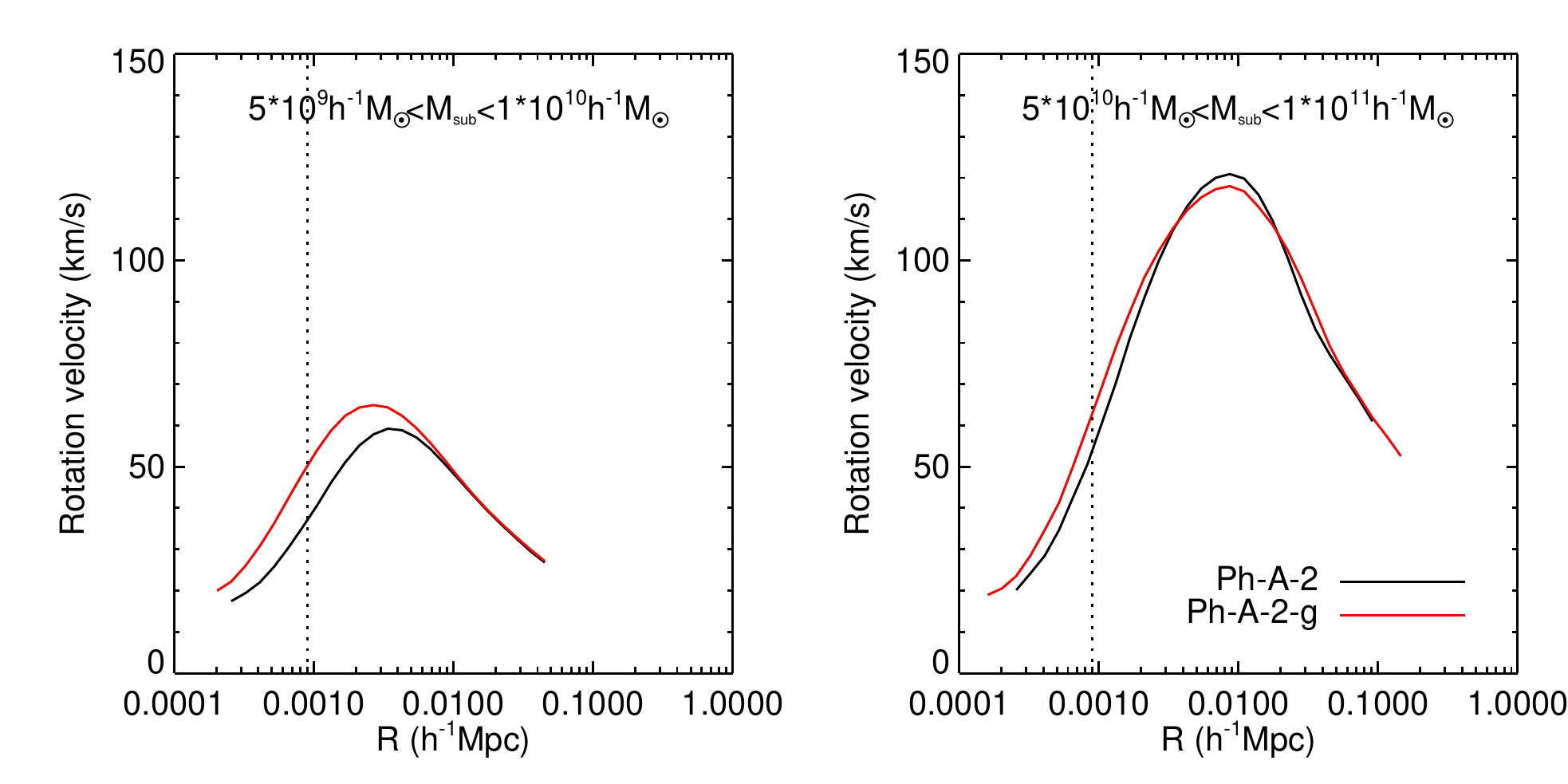}
    \caption{The rotation curves of subhaloes in two mass bins: $[5\times 10^9,10^{10}]h^{-1}M_{\odot}$ (left) and $[5\times 10^{10},10^{11}]h^{-1}M_{\odot}$ (right). We stack 20 subhaloes within viral radius for each bin. Black and red lines correspond to the Ph-A-2 and the Ph-A-2-g subhaloes respectively. Vertical dotted line represents 2.8 times softening length in the Ph-A-2 simulation.}

	\label{fig:stack_rotation_curve}
	\end{figure*}
    
	In order to see more clearly the difference in the internal structure of subhaloes between the DMO and NHD simulation, we select two 20 random low- and high-mass subhalo samples from both Ph-A-2 and Ph-A-2-g simulations. The mass range of the low-mass sample is $[5\times 10^9,10^{10}]h^{-1}M_{\odot}$, and it is $[5\times 10^{10},10^{11}]h^{-1}M_{\odot}$ for the high-mass sample. Then we stack rotational curves of all subhaloes in each sample and display them in Fig.~\ref{fig:stack_rotation_curve}, the vertical dotted lines indicate 2.8 times softenning length at which the force calculation in the Gadget code becomes Newtonian. While the outer parts of rotation curves are quite similar, differences in the inner part are noticeable between the Ph-A-2 and Ph-A-2-g. In particularly for the low mass subhalo samples, inner density is substantially higher for the subhaloes in the Ph-A-2-g than Ph-A-2 simulation. For the high mass subhaloes sample, even its  $v_{max}$ is slightly lower in the Ph-A-2-g than Ph-A-2, the inner density is still slightly higher in the Ph-A-2-g. In the Appendix, we provide one-to-one comparisons of rotational curves of 4 randomly selected matched individual subhaloes in both DMO and NHD runs, results are consistent with the statistical results presented here.

    \begin{figure*}[htb]
	\includegraphics[width=\textwidth]{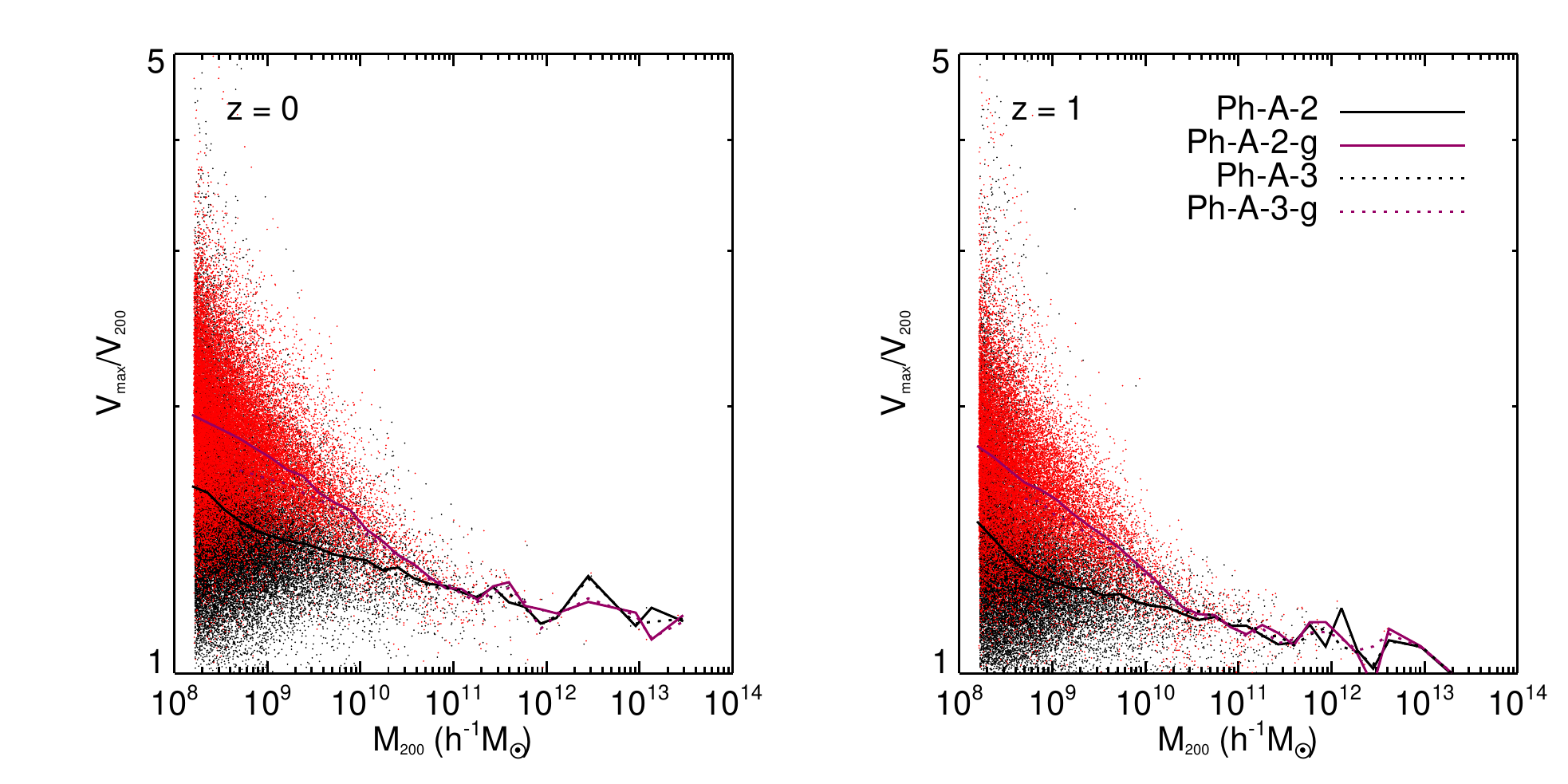}
	\caption{$V_{max}/V_{200} - M_{200}$ relations of the field halos at redshift $z=0$ (left panel) and $z=1$ (right panel) in the DMO and NHD simulations. The median values of the level 2 (solid lines) and 3 (dashed lines) are shown. The DMO and NDH runs are distinguised with different colors as shown in the label of the right panel.}
	\label{fig:nomatch_iso_m_vs_vmax}
	\end{figure*}

Is the difference in the $V_{max}$--$M_{sub}$ relation of subhalo between the DMO and NHD simulations due to interactions of subhaloes with their host halo? We trace the progenitors of the present-day subhaloes just before accretion in both DMO and NHR simulations, and plot their $V_{max}$ as a function of the progenitor mass in the right panel of Fig.~\ref{fig:m_vs_vmax_z0}. Apparently, the relation for the progenitors is quite similar to that of subhaloes, namely low-mass progenitors are statistically more concentrated in the NHD than in  DMO simulations. As the accretion time distribution of the progenitors is quite broad,  the low-mass progenitors are more concentrated in our NHD than in DMO simulations over a quite broad redshift range, and thus the different subhalo $V_{max}$--$M_{sub}$ relation between the DMO and NHD simulations simply reflects the different structural properties between two kinds of simulations.
   
Indeed, previous studies suggest that, compared to the DMO simulation, dark matter haloes are slightly more concentrated in hydrodynamic simulation (\citealt{Rasia2004} and \citealt{Lin2006}). For instance, \cite{Lin2006} found that about $3-8$ percent difference in halo concentration parameter $c$ between the NHD and DMO simulations  for haloes with virial mass from $10^{13}$ to $7 \times 10^{14} h^-1M_{\odot}$.  In Fig.~\ref{fig:nomatch_iso_m_vs_vmax}, we examine the concentration-mass relation of isolated dark matter haloes with our own simulations and extend to much lower halo mass. Here the isolated haloes are selected within a cluster-centric radius $[1.5, 3]R_{200} $, and concentrated parameters of dark matter haloes are represented with  $V_{max}/V_{200}$. Results for two different resolutions and two redshifts are shown. Black and red dots show the results for Ph-A-2 and Ph-A-2-g respectively, the different lines show median values of $V_{max}/V_{200}$ in each halo mass bin. Clearly, concentration parameter is larger for low-mass haloes with masses less massive than $2 \times 10^{10}h^{-1}M_{\odot}$ in the NHD than DMO simulations. Above this mass, the concentration parameter of dark matter haloes are similar between two sets of simulations, slightly inconsistent with previous studies such as \cite{Lin2006}.  This inconsistency is mainly due to the different derivation of the concentration parameter.  \cite{Lin2006} derived the  concentration parameter of a dark matter halo by fitting its density profile. When we fit $c$ rather than using the proxy $V_{max}/V_{200}$, our result is consistent with \cite{Lin2006}.

\subsection{THE EVOLUTION OF SUBHALOES}
\label{sec:ev}

\begin{figure*}
	\includegraphics[width=\textwidth]{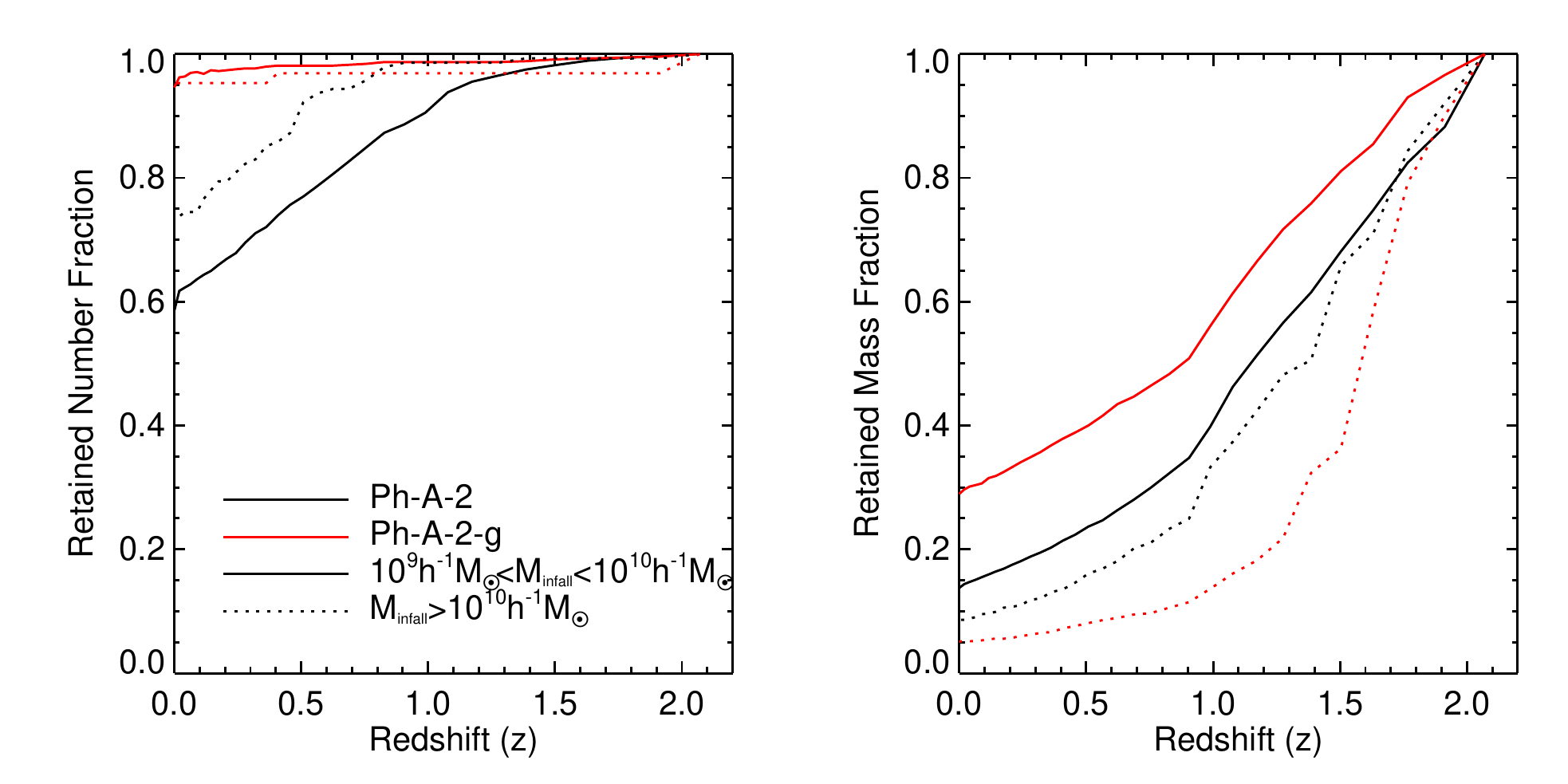}
	\caption{The survived number (left panel) and mass (right panel) fraction of the progenitors accreted at redshift $z=2$ as a function of redshift. Results for the Ph-A-2 and Ph-A-2-g are distinguished with different colors, and results for two different progenitor mass samples are distinguished with different line styles as indicated in the label.}
	\label{fig:servive_mass_fraction}
	\end{figure*}

In this subsection, we compare the evolution of subhaloes once they were accreted into the Ph-A halo with our matched highest resolution simulations--the Ph-A-2 and Ph-A-2-g. Following \cite{Xie2015}, in Fig.~\ref{fig:servive_mass_fraction} we present the survival number and retained mass fraction of the progenitors accreted at a fixed epoch $z=2$ as a function of redshift. Here we consider two progenitor samples according to their halo mass, $M_{infall} > 10^{10}h^{-1}M_{\odot}$ and $10^9h^{-1}M_{\odot} < M_{infall} < 10^{10}h^{-1}M_{\odot}$. Apparently, the survival number fraction is larger in the Ph-A-2-g  in relative to the Ph-A-2 simulation.  For relatively more massive progenitors, more than 90 percent of them survive to the present day in the Ph-A-2-g, while only 75 percent of them survive in the Ph-A-2. For low-mass progenitors, the survival fraction is quite similar to their high-mass counterparts in the Ph-A-2-g run, while the fraction decreases to 60 percent in the Ph-A-2 simulation. This is mainly a consequence of higher concentration of low-mass dark matter haloes in the NHD than DMO simulations. Note that the survival number fraction of subhaloes in our Ph-A-2-g simulation is consistent with that of hydrodynamic simulation with cooling and star formation of \cite{Bahe2019}.

The right panel of the Fig.~\ref{fig:servive_mass_fraction} shows the evolution of the retained mass fraction of the same progenitor samples as a function of redshift. Here the retained mass fraction is defined as the ratio of the total retained mass to the total progenitor mass at each redshift. For relatively massive progenitors, their mass-loss is more rapid in the Ph-A-2-g than Ph-A-2 simulation, at intermediate redshift, say $z=1$, the retained mass fraction is about 3 times higher in the Ph-A-2 than Ph-A-2-g simulation. This is expected because subhaloes suffer from additional ram-pressure stripping in the hydrodynamic simulation.  For the low-mass progenitors, the relation is reverse, the mass-loss of subhaloes is less efficient in the Ph-A-2-g than Ph-A-2. As low-mass haloes have higher concentration in NHD simulation, which counteracts the effect of ram-pressure and results in less efficient mass stripping.  These subhalo mass dependence survival/retained fractions should largely account for the different shapes in the subhalo mass/$V_{max}$ functions between the DMO  and NHD simulations.

\section{CONCLUSION}
\label{sec:con}
We perform a set of matched hydrodynamic simulations of a rich cluster-sized dark matter halo from the Phoenix Project, and compare in detail the subhalo population in the halo simulated with and without non-radiative hydrodynamics, and with 3 different numerical resolutions. Our results can be summarized as follows.

1) The subhalo mass function of the cluster-sized dark matter halo has different shapes between the DMO and NHD simulations. The cumulative subhalo mass function can be approximated with $N(M) \propto M^{-1}$  for the DMO simulation, while it is $N(M) \propto M^{-1.1}$ for the NHD run. Subhaloes are more abundant in the DMO runs for subhaloes more massive than $2 \times 10^{9}h^{-1}M_{\odot}$, below this mass, there are more subhaloes in the NHD. Similarly, the subhalo $V_{max}$ function is also steeper ($n \sim -3.4$ versus $-4.4$) in the NHD hydrodynamic than in the DMO simulation. Above $V_{max} \sim 70 kms^{-1}$, subhaloes are more abundant in the DMO simulation, while below this value there are more subhaloes in the NHD simulation.

 2) For subhaloes less massive than $10^{10}h^{-1}M_{\odot}$, subhaloes have larger concentration in the NHD than DMO simulation, above this mass subhaloes have similar concentration in two sets of simulations. We show that this is mainly because that progenitors of the present-day subhaloes are more concentrated before accretion in the NHD than in the DMO simulation.  At $z=0$ and $z=1$, the low-mass field dark matter haloes have larger concentration parameters in the NHD than DMO simulation.

3) The relatively larger concentration parameters of the low mass field haloes also result in larger survival number and larger retained mass fraction of the low mass progenitors in the NHD than DMO simulation. 

Based on above results, a simple picture to interpret the differences in the subhalo population between the DMO and NHD simulation can be summarized as follows. Difference in structural properties of dark matter haloes varies with halo mass between the DMO and NHD simulations, with the difference being large with the decreasing halo mass. As a result, when these dark matter halo merge into their primary halo, their survival ability is larger in the NHD than DMO simulation, leading to the more abundant low mass subhaloes in the NHD simulation.

Note that our main results are based on a set of simulations of a single cluster-sized dark matter haloes, statistical results are necessary and will be presented in a future work.

\section*{Acknowledgements}
We are grateful to useful discussions with Dr. Shihong Liao.
This work is supported by the National Key R\&D Program of China (NO. 2017YFB0203300), and the Key Program of the National Natural Science Foundation of China (NFSC) through grant 11733010.
	





\appendix{Comparisons of Matched Subhaloes}
\section{Matching subhaloes}

As we will make one-to-one comparison of subhaloes in the DMO and NHD simulations here, we follow \citet{Schaller2015} to match subhaloes according to their particle Ids. In practise, we select subhaloes with more than 200 dark matter particles in each level NHD simulation. For each of them, we identify 50 most bound dark matter particles. If a subhalo in its DMO counterpart contains 25 of them, we consider the subhalo a potential candidate for a matched one in the DMO run. Now we further require the subhalo in the NHD simulation also containing half of 50 most bound dark matter of the candidate. If both conditions are satisfied, we consider them a matched subhalo pair. For our level 2 simulations, more than $53.2\%$ subhaloes with $M_{sub} > 10^{9}h^{-1}M_{\odot}$  can be matched at z = 0, and $81.87\%$ for subhaloes above $10^{10}h^{-1}M_{\odot}$, in consistent with \citet{Schaller2015}.

\begin{figure*}
	\includegraphics[width=\textwidth]{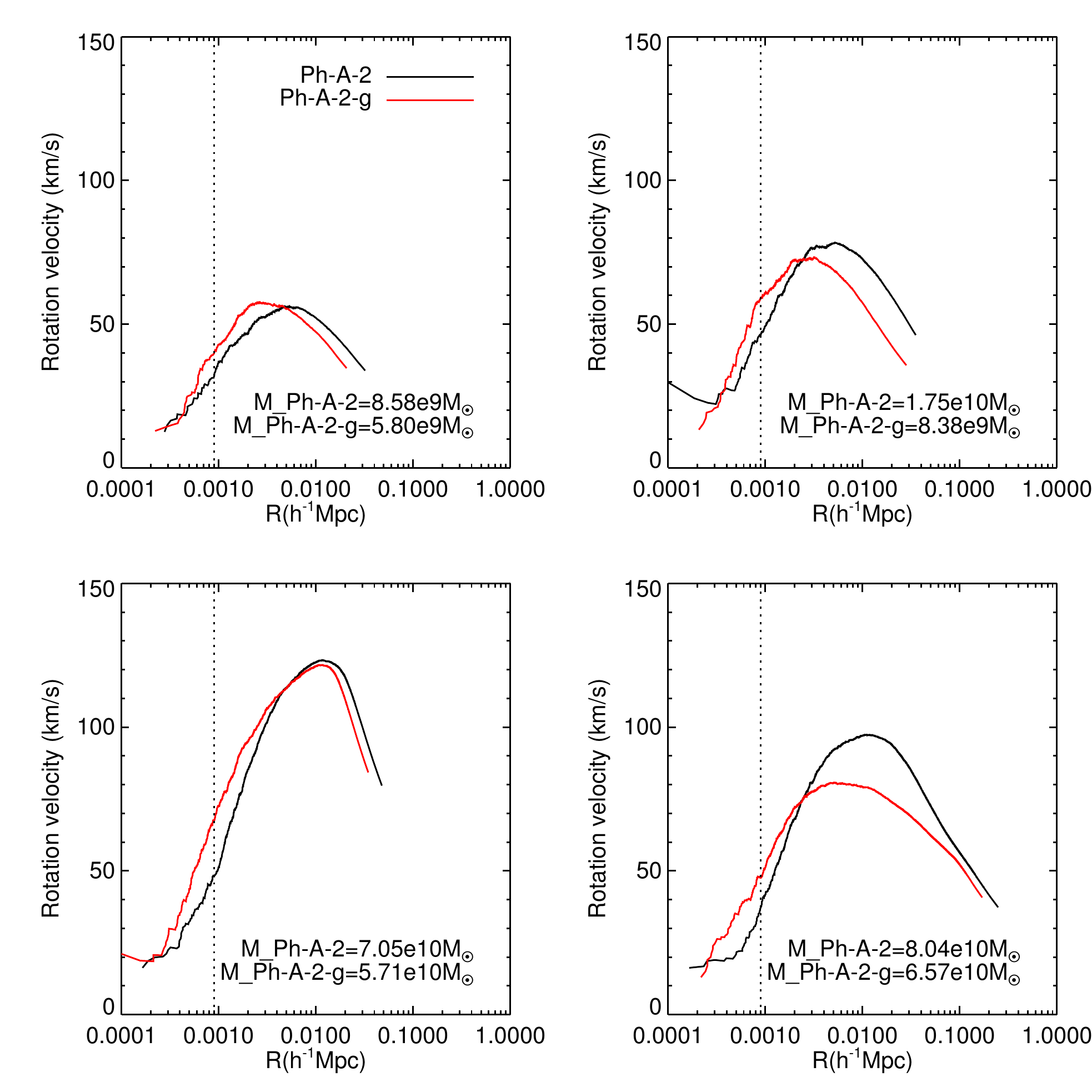}
	\caption{Rotation curve for 4 pairs of 4 random matched subhalos. The masses of these subhaloes are indicated in the label.}
	\label{fig:matched_rotation_curve}
\end{figure*}

In Fig.~\ref{fig:matched_rotation_curve}, we select 4 random pairs of matched subhalos in $[5\times 10^9,10^{10}]h^{-1}M_{\odot}$ and $[5\times 10^{10},10^{11}]h^{-1}M_{\odot}$ subhalo samples, and plot their rotation curves. The overall results are consistent with the statistical ones presented in the figure 4, namely subhaloes are more centrally concentrated in the NHD than DMO runs. It is also quite interesting that, even for subhaloes with even larger $V_{max}$ in the DMO run, central density is still higher in the NHD run. This should account for the fact that the survival number fraction is higher in the NHD than DMO runs even they have similar concentration.

 \begin{figure}
  \centering
	\includegraphics[width=0.5\textwidth]{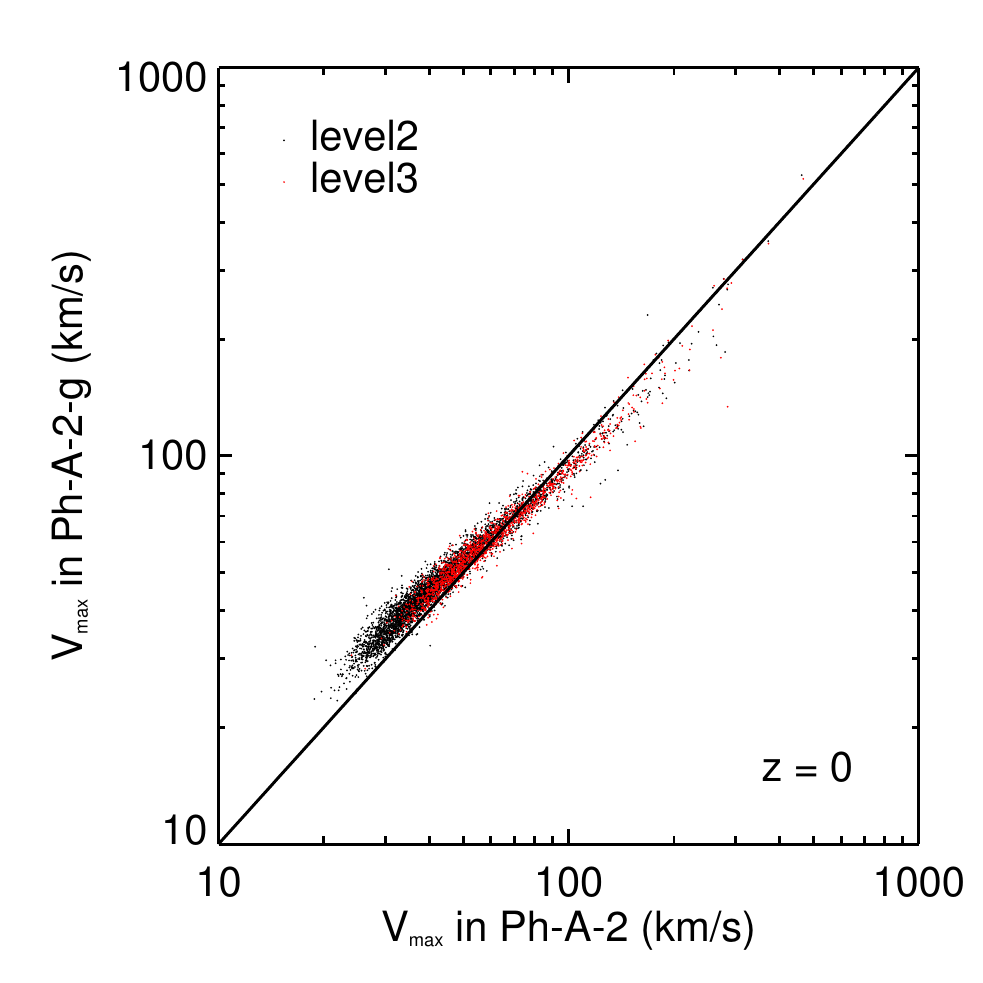}
	\caption{$V_{max}$ of matched subhaloes at $z=0$. Each point in the figure represents a matched subhalo in DMO and NHD simulations. Diagonal shows the case when subhalo has the same $V_{max}$ in both simulations. Black and red points here are subhaloes from level 2 and level 3 simulations respectively.}
	\label{fig:vmax_z0}
\end{figure}

In figure~\ref{fig:vmax_z0}, we compare $V_{max}$ of all our matched subhalo samples. Clearly, subhaloes are more concentrated in the NHD simulations, in particularly the low mass subhaloes, consistent with the results presetned in the Fgiure ~\ref{fig:m_vs_vmax_z0}.

\bibliographystyle{mn2e}
\setlength{\bibhang}{2.0em}
\setlength\labelwidth{0.0em}
\bibliography{pha-4}

\begin{thebibliography}{99}

\bibitem[\protect\citeauthoryear{Bah\'e et al.}{2019}]{Bahe2019}
Bah\'e Y. M., et al., 2019, MNRAS, 485, 2287

\bibitem[\protect\citeauthoryear{Chua et al.}{2017}]{Chua2017}
Chua K. T. E., Pillepich A., Rodriguez-Gomez V., Vogelsberger M., Bird S., Hernquist L., 2017, MNRAS, 472, 4343

\bibitem[\protect\citeauthoryear{Davis et al.}{1985}]{Davis1985}
Davis M., Efstathiou G., Frenk C. S., White S. D. M., 1985, ApJ, 292, 371

\bibitem[\protect\citeauthoryear{De Lucia et al.}{2004}]{De Lucia2004}
De Lucia G., Kauffmann G., Springel V., White S. D. M., Lanzoni B., Stoehr F., Tormen G., Yoshida N., 2004, MNRAS, 348, 333

\bibitem[\protect\citeauthoryear{Despali et al.}{2017}]{Despali2017}
Despali G., Vegetti S., 2017, MNRAS, 469, 1997

\bibitem[\protect\citeauthoryear{Diemand et al.}{2004}]{Diemand2004}
Diemand J., Moore B., Stadel J., 2004, MNRAS, 352, 535

\bibitem[\protect\citeauthoryear{Diemand et al.}{2007}]{Diemand2007}
Diemand J., Kuhlen M., Madau P., 2007, ApJ, 667, 859

\bibitem[\protect\citeauthoryear{Dolag et al.}{2009}]{Dolag2009}
Dolag K., Borgani S., Murante G., Springel V., 2009, MNRAS, 399, 497

\bibitem[\protect\citeauthoryear{Duffy et al.}{2010}]{Duffy2010}
Duffy A. R., Schaye J., Kay S. T., Dalla Vecchia C., Battye R. A., Booth C. M., 2010, MNRAS, 405, 2161

\bibitem[\protect\citeauthoryear{Elahi et al.}{2016}]{Elahi2016}
Elahi P. J., et al., 2016, MNRAS, 458, 1096

\bibitem[\protect\citeauthoryear{Fang et al.}{2009}]{Fang2009}
Fang, T., Humphrey, P., \& Buote, D. 2009, ApJ, 691, 1648

\bibitem[\protect\citeauthoryear{Gao et al.}{2004}]{Gao2004}
Gao L., De Lucia G., White S. D. M., Jenkins A., 2004, MNRAS, 352, L1

\bibitem[\protect\citeauthoryear{Gao et al.}{2012}]{Gao2012}
Gao L., Navarro J. F., Frenk C. S., Jenkins A., Springel V., White S. D. M., 2012, MNRAS, 425, 2169

\bibitem[\protect\citeauthoryear{Garrison-Kimmel et al.}{2017}]{Garrison-Kimmel2017}
Garrison-Kimmel, S., Wetzel, A., Bullock, J. S., et al. 2017, MNRAS, 471, 1709

\bibitem[\protect\citeauthoryear{Ghigna et al.}{1998}]{Ghigna1998}
Ghigna S., Moore B., Governato F., Stadel J., 1998, MNRAS, 300, 146

\bibitem[\protect\citeauthoryear{Ghigna et al.}{2000}]{Ghigna2000}
Ghigna S., Moore B., Governato F., Lake G., Quinn T., Stadel J., 2000, ApJ, 544, 616

\bibitem[\protect\citeauthoryear{Graus et al.}{2018}]{Graus2018}
Graus A. S., Bullock J. S., Boylan-Kolchin M., Nierenberg A. M., 2018, MNRAS, 480, 1322

\bibitem[\protect\citeauthoryear{Han et al.}{2016}]{Han2016}
Han, J., Cole, S., Frenk, C. S., \& Jing, Y. 2016, MNRAS, 457, 1208

\bibitem[\protect\citeauthoryear{Han et al.}{2018}]{Han2018}
Han J., Cole S., Frenk C. S., Benitez-Llambay A., Helly J., 2018, MNRAS, 474, 604

\bibitem[\protect\citeauthoryear{Jiang et al.}{2014}]{Jiang2014}
Jiang, L., Helly, J. C., Cole, S., \& Frenk, C. S. 2014, MNRAS, 440, 2115

\bibitem[\protect\citeauthoryear{Klypin et al.}{1999}]{Klypin1999}
Klypin A., Kravtsov A. V., Valenzuela O., Prada F., 1999, ApJ, 522, 82

\bibitem[\protect\citeauthoryear{Kravtsov et al.}{2004}]{Kravtsov2004}
Kravtsov A. V., Gnedin O. Y., Klypin A. A., 2004, ApJ, 609, 482

\bibitem[\protect\citeauthoryear{Kuhlen et al.}{2007}]{Kuhlen2007}
Kuhlen M., Diemand J., Madau P., 2007, ApJ, 671, 1135

\bibitem[\protect\citeauthoryear{Libeskind et al.}{2010}]{Libeskind2010}
Libeskind N. I., Yepes G., Knebe A., Gottlober S., Hoffman Y., Knollmann S. R., 2010, MNRAS, 401, 1889 

\bibitem[\protect\citeauthoryear{Lin et al.}{2006}]{Lin2006}
Lin W. P., Jing Y. P., Mao S., Gao L., McCarthy I. G., 2006, ApJ, 651, 636

\bibitem[\protect\citeauthoryear{Macci\'o et al.}{2006}]{Maccio2006}
Macci\'o A. V., Moore B., Stadel J., Diemand J., 2006, MNRAS, 366, 1529

\bibitem[\protect\citeauthoryear{Moore et al.}{1998}]{Moore1998}
Moore, B., Lake, G., \& Katz, N. 1998, ApJ, 495, 139

\bibitem[\protect\citeauthoryear{Moore et al.}{1999}]{Moore1999}
Moore B., Ghigna S., Governato F., Lake G., Quinn T., Stadel J., Tozzi P., 1999 ApJ, 524, 19

\bibitem[\protect\citeauthoryear{Nagai et al.}{2005}]{Nagai2005}
Nagai D., Kravtsov A. V., 2005, ApJ, 618, 557

\bibitem[\protect\citeauthoryear{Puchwein et al.}{2005}]{Puchwein2005}
Puchwein E., Bartelmann M., Dolag K., Meneghetti M., 2005, A\&A, 442, 405

\bibitem[\protect\citeauthoryear{Rasia et al.}{2004}]{Rasia2004}
Rasia, E., Tormen, G., \& Moscardini, L. 2004, MNRAS, 351, 237

\bibitem[\protect\citeauthoryear{Richings et al.}{2020}]{Richings2020}
Richings J., Frenk C., Jenkins A., Robertson A., 2020, MNRAS, 492.5780R

\bibitem[\protect\citeauthoryear{Romano-D\'iaz et al.}{2010}]{Romano-Diaz2010}
Romano-D\'iaz, E., Shlosman, I., Heller, C., \& Hoffman, Y. 2010, ApJ, 716, 1095

\bibitem[\protect\citeauthoryear{Sawala et al.}{2013}]{Sawala2013}
Sawala T., Frenk C. S., Crain R. A., Jenkins A., Schaye J., Theuns T., Zavala J., 2013, MNRAS, 431, 1366

\bibitem[\protect\citeauthoryear{Sawala et al.}{2017}]{Sawala2017}
Sawala T., Pihajoki P., Johansson P. H., Frenk C. S., Navarro J. F., Oman K. A., White S. D. M., 2017, MNRAS, 467, 4383

\bibitem[\protect\citeauthoryear{Schaller et al.}{2015}]{Schaller2015}
Schaller, M., Frenk, C. S., Bower, R. G., et al. 2015, MNRAS, 451, 1247

\bibitem[\protect\citeauthoryear{Springel et al.}{2001}]{Springel2001a}
Springel V., White S. D. M., Tormen G., Kauffmann G., 2001, MNRAS, 328, 726

\bibitem[\protect\citeauthoryear{Springel et al.}{2001}]{Springel2001b}
Springel V., Yoshida N., White S. D. M., 2001, NewA, 6, 79

\bibitem[\protect\citeauthoryear{Springel et al.}{2005}]{Springel2005}
Springel V., White S. D. M., Jenkins A., et al., 2005, Nat, 435, 629

\bibitem[\protect\citeauthoryear{Stoehr et al.}{2002}]{Stoehr2002}
Stoehr F., White S. D. M., Tormen G., Springel V., 2002, MNRAS, 335, L84

\bibitem[\protect\citeauthoryear{Stoehr et al.}{2003}]{Stoehr2003}
Stoehr F., White S. D. M., Springel V., Tormen G., Yoshida N., 2003, MNRAS, 345, 1313

\bibitem[\protect\citeauthoryear{Tormen et al.}{2004}]{Tormen2004}
Tormen G., Moscardini L., Yoshida N., 2004, MNRAS, 350, 1397

\bibitem[\protect\citeauthoryear{van den Bosch et al.}{2018}]{vandenBosh2018}
van den Bosch F. C., Ogiya G., 2018, Mon. Not. R. Astron. Soc., 475, 4066

\bibitem[\protect\citeauthoryear{Weinberg et al.}{2008}]{Weinberg2008}
Weinberg D. H., Colombi S., Dav\'e R., Katz N., 2008, ApJ, 678, 6

\bibitem[\protect\citeauthoryear{Xie \& Gao}{2015}]{Xie2015}
Xie, L., \& Gao, L. 2015, MNRAS, 454, 1697

\bibitem[\protect\citeauthoryear{Zhu et al.}{2016}]{Zhu2016}
Zhu Q., Marinacci F., Maji M., Li Y., Springel V., Hernquist L., 2016, MNRAS, 458, 1559





\bibitem[\protect\citeauthoryear{}{}]{}


\bibitem[\protect\citeauthoryear{}{}]{}


\bibitem[\protect\citeauthoryear{}{}]{}


\bibitem[\protect\citeauthoryear{}{}]{}


\bibitem[\protect\citeauthoryear{}{}]{}


\bibitem[\protect\citeauthoryear{}{}]{}


\bibitem[\protect\citeauthoryear{}{}]{}



\end{thebibliography}
\label{lastpage}
\end{document}